\documentclass[onecolumn, prl,
 amsmath,amssymb,
 superscriptaddress,aps,
]{revtex4-2}
\usepackage[utf8]{inputenc}
\usepackage[english]{babel}
\usepackage{amsmath}
\usepackage{hyperref}
\usepackage{amsfonts}
\usepackage{amssymb}
\usepackage{graphicx}
\usepackage{epstopdf}
\usepackage{xcolor}
\usepackage[dvipsnames]{xcolor}
\usepackage{ulem}
\usepackage{setspace}

\begin{document}

\title{Evolution of dense liquid marbles under light tapping}

\author{Amaelle Sprocq }
\affiliation{Université de Lille, CNRS, UMR 8520-IEMN, Lille, France.}

\author{Théo Dhoudain}
\affiliation{Université de Lille, CNRS, UMR 8520-IEMN, Lille, France.}

\author{Alexis Duchesne}
\affiliation{Université de Lille, CNRS, UMR 8520-IEMN, Lille, France.}

\author{Guillaume Lagubeau}\email{guillaume.lagubeau@univ-lille.fr}
\affiliation{Université de Lille, CNRS, UMR 8520-IEMN, Lille, France.}

\begin{abstract}
Liquid marbles interface is composed by a dense network of grains trapped by capillarity at their surface. The mix of capillarity and granular character make their interfaces remarkable, allowing non wetting, high mobility and strong robustness which are a basis for numerous applications.  
A canonical experiment showing out of equilibrium nature of granular material is the so called tapping experiment, where a dry granular medium under vertical vibration reorganize an evolve toward a denser final state function of the vibration amplitude
In this paper we explore the out of equilibrium nature of densely covered liquid marbles by measuring their shape, compaction and stresses evolution under vertical vibration.  We show that liquid marble do indeed reorganize when submitted to taps, leading to higher internal pressure and lower effective surface tension.

\end{abstract}

\maketitle


Liquid marbles are liquids coated by a shell formed by a dense network of submillimetric grains trapped by capillarity \cite{mchale2011liquid,bormashenko2017liquid,protiere2023particle,tenjimbayashi2023liquid}. Such object exhibit remarkable mobility\cite{despointes2026ultra} and robustness \cite{sneha2024liquid} making them useful in numerous applications \cite{bormashenko2017liquid,kato2023surface,jin2023marbles,tong2025advancements,jampani2026packaging} such as digital microfluidics, oil encapsulation, biochemical analysis, materials synthesis, microsensors, environmental monitoring, soft robotics, or particle separation. Their mechanics was first understood in analogy to non wetting liquid droplets possessing effective capillarity and viscosity \cite{aussillous2001liquid,aussillous2006properties,celestini2018propulsion}. However, the dense network of grains present at their interface leads unique properties such as a strong dependence of shell stresses and internal dissipation on both the fraction of the shell covered by grains, and its preparation history \cite{lagubeau2014armoring,takai2026static}. It has then be possible to connect armored interface wrinkling to jamming \cite{lagubeau2014armoring}, and their mobility threshold to friction between grains \cite{takai2026static}. 
Granular effects have been studied extensively in the case of flat armored interfaces called "particle rafts" \cite{binks2005naturally,cicuta2009granular,jambon2017wrinkles,planchette2025unjamming}, but the granular properties of the shell remain vastly unexplored in the case of liquid marbles. A canonical experiment for granular materials \cite{andreotti2013granular} is tapping: under vertical vibration, a granular medium experience compaction reaching a final state which depends on the vibration intensity\cite{mehta1990phenomenological,knight1995density,philippe2002compaction}. This phenomena originates from the out-of equilibrium nature of granular material: dissipation trap them in a metastable state, and by injecting energy, the material can reorganize, explor its possible configurations and find an optimum. In this paper we aim at exploring the out of equilibrium nature of liquid marble densely covered by particles by measuring shape, compaction and stresses evolution under vertical loading analogue to tapping.

\begin{figure*}[]
	\centering
	\includegraphics[scale=0.29]{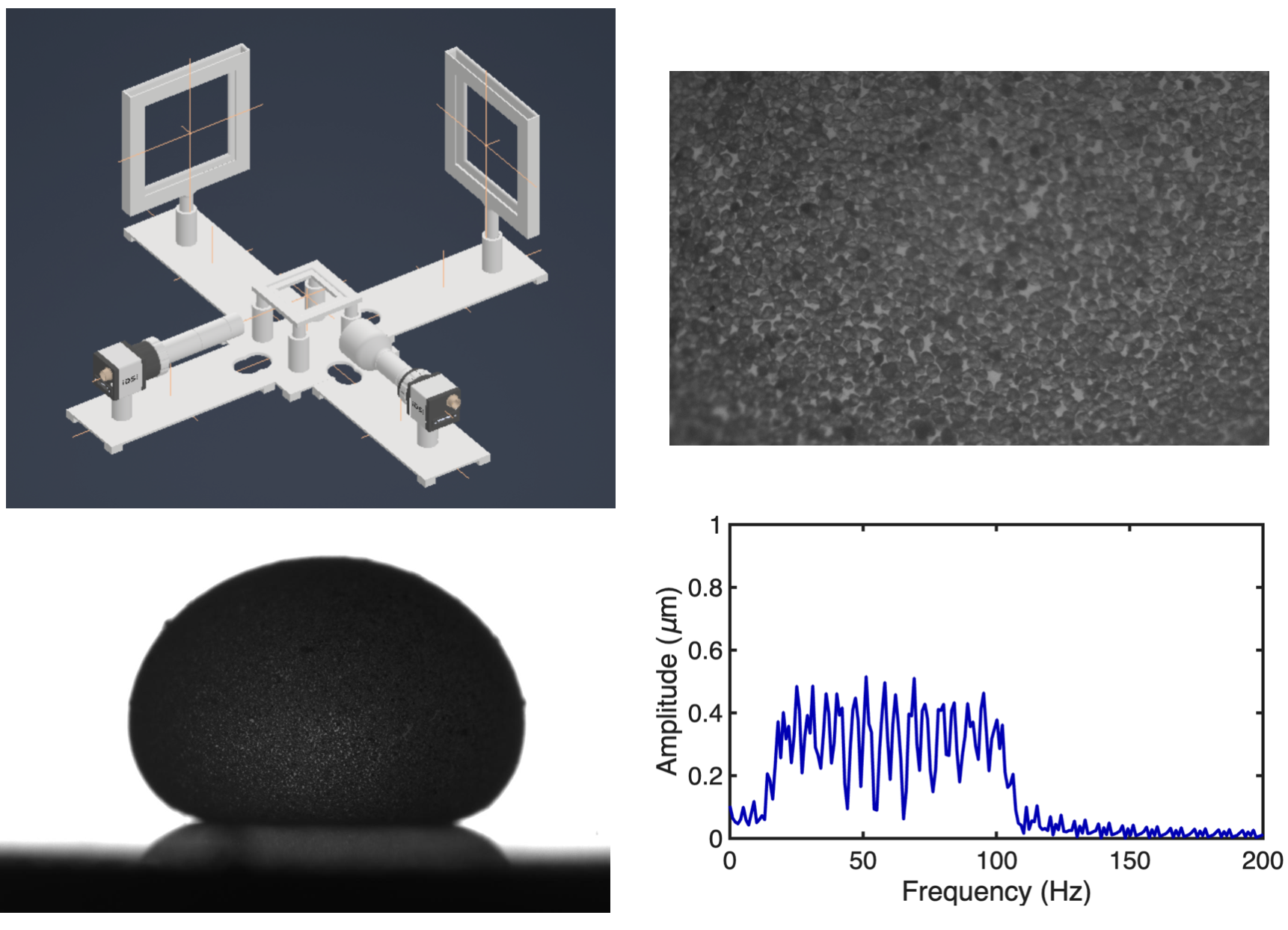}
	\caption{Experimental Setup. a. overview of the experimental setup: A drop liquid marble covered by lycopodium powder is deposited on a piezoelectric speaker and imaged by two cameras, under back-lighting. b. Side view of a $40\mu L$ water droplet covered by lycopodium powder. c. Details of the liquide marble surface (around 1\%). The typical size of a grain is $ \mu m$ d. Signal emitted by the piezoelectric speaker (Amplitude vs Frequency).}
	\label{Fig1}
\end{figure*}

A de-ionized water drop of volume $20\mu L$ or $40\mu L$ covered by lycopodium powder is deposited on a piezoelectric speaker (Fig.~\ref{Fig1} a.) and imaged by two cameras, under back-lighting. The first camera images the full profile (Fig.~\ref{Fig1} b.), allowing to measure drop volume Area, and shell stresses. The second camera has a much higher magnification, imaging around 1\% of the droplet surface, allowing to view the grain network (Fig.~\ref{Fig1} c.), and measures surface fraction. The drop is vibrated vertically using a short burst ("tap") of large band small amplitude vibration (Fig.~\ref{Fig1} d.).
Profile and local network photos are taken just before tapping, and this process is repeated sequentially with a delay between two taps, of $1.15s$, allowing for both relaxation of the droplet and photography.

\begin{figure*}[!h]
	\centering
	\includegraphics[scale=0.29]{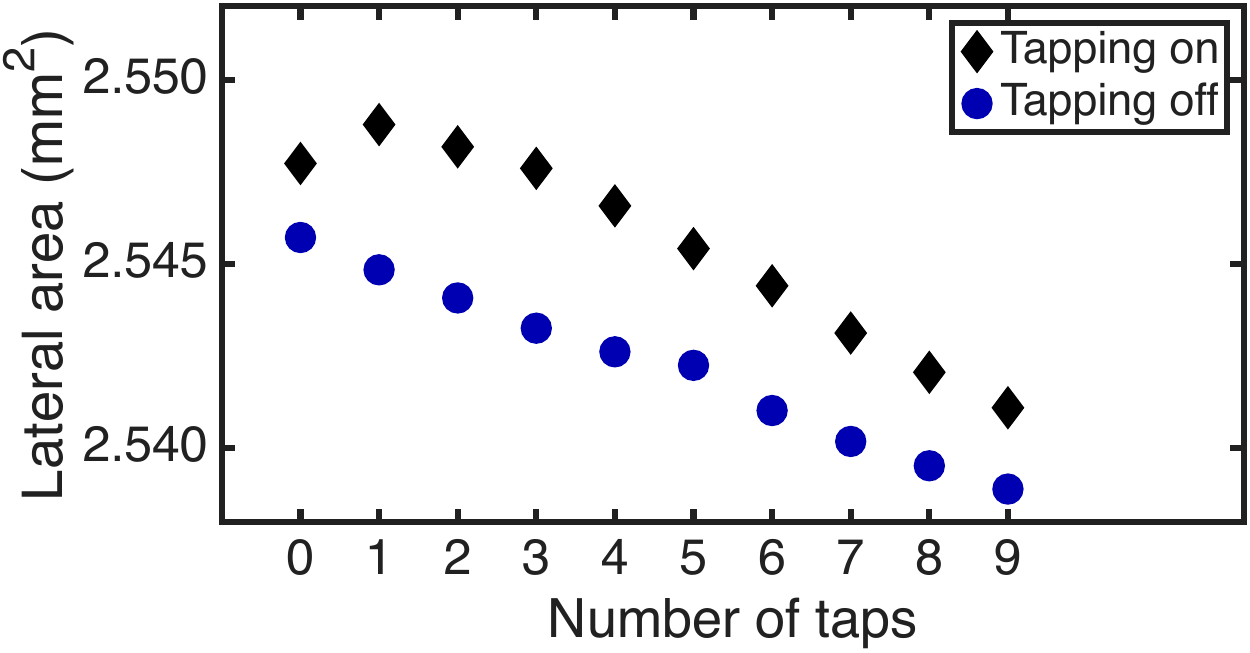}
	\caption{Lateral Area of a $20 \mu L$ liquid marble covered by lycopodium with a surface fraction of $83\%$. Comparison of area evolution after 9 successive taps (black losanges) with reference experiment were tapping are turned off (blue circles) with   }
	\label{VsRef}
\end{figure*}

We first compare a $20\mu L$ drop lateral area evolution when submitted to taps to a reference drop not submitted to taps (Fig.~\ref{VsRef}), in the same imaging setup. The lateral area of the reference droplet is monotonously slowly reduced by evaporation. In contrast, under tapping, lateral area is increased by the first tap. This increase is not observed for later taps, for which evaporation is dominant. 

\begin{figure*}[!h]
	\centering
	\includegraphics[scale=0.29]{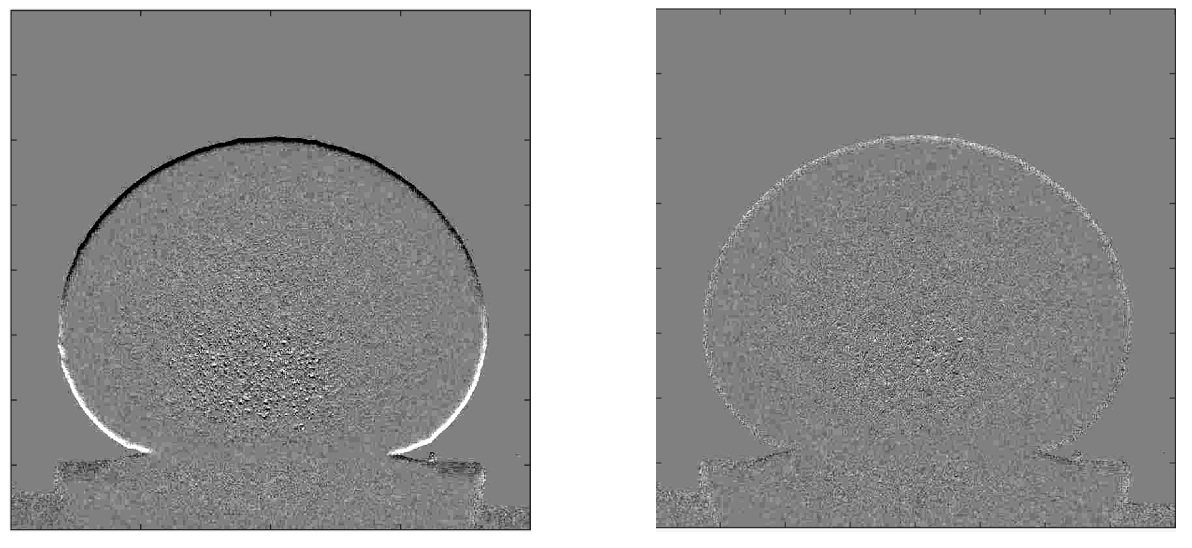}
	\caption{Liquid marble of $20 \mu L$ submitted to sequential tap  a. (left) difference before and after the first tap. b. (right) difference before and after the third tap. same grey scale}
	\label{DiffIm}
\end{figure*}

The lateral increase after the first tap coincide with a shape change of the droplet evidenced in Fig.\ref{DiffIm} a. where the difference of profile image between before and after the first tap is shown. We observe in the image difference a darker zone of the top and a brighter zone at the bottom, evidence of a smaller contact area as well as a higher. In can be seen as the result of the drop inflating, not in the sense of a volume change, but in the sense of an internal pressure increase. In comparison, the image difference between before and after the third tap do not exhibit the same shape change, and present only a smaller effect of volume reduction due to evaporation (Fig. \ref{DiffIm} b.).

To isolate and quantitatively investigate the tapping effect, we repeated the experiment, keeping the tapping the same, for two distinct drop volumes ($20 \mu L$ and $40 \mu L$) using a drop preparation process devised to create close initial properties (see Appendix A). In table \ref{TableDataMarbles} we present the characteristics of the liquid marbles used in this study (Volume Surface Fraction $\Phi$, effective surface tension $\gamma$, principal stress asymmetry $\delta \gamma$). The drops are densely packed, with a surface fraction close to, but $1\%$ lower to where wrinkling is observed. For drops of $20 \mu L$, effective tension and anisotropy are consistent with \cite{lagubeau2014armoring} as well as surface fraction at wrinkling although slightly higher ($83.7 \%$ compared to $82.8\%$).

\begin{table}[!h]
\begin{center}
\caption{Summary of liquid marbles characteristics used in this study. Dispersion of data quantified by standard deviation.}
	\label{Table1}
\begin{tabular}{ |c|c|c|c|c|c| } 
 \hline
 
 Data set & Volume ($\mu L$) & Surface Fraction (\%) & $\gamma (mN.m^-1)$ & $\delta \gamma (mN.m^-1)$ &Number of experiments \\ 
 1 & $20.01 \pm 0.65$ & $82.9 \pm 1.7$ & $42.3 \pm 3.3$ &  $1.98\pm0.80$ & 35\\ 
 2 & $39.9 \pm 1.6$ & $83.6\pm 1.5$ & $41.0 \pm 2.4$ &  $3.08 \pm 0.92$ & 28\\ 
 \hline
\end{tabular}
\end{center}
\label{TableDataMarbles}
\end{table}

\begin{figure*}[!h]
	\centering
	\includegraphics[scale=0.29]{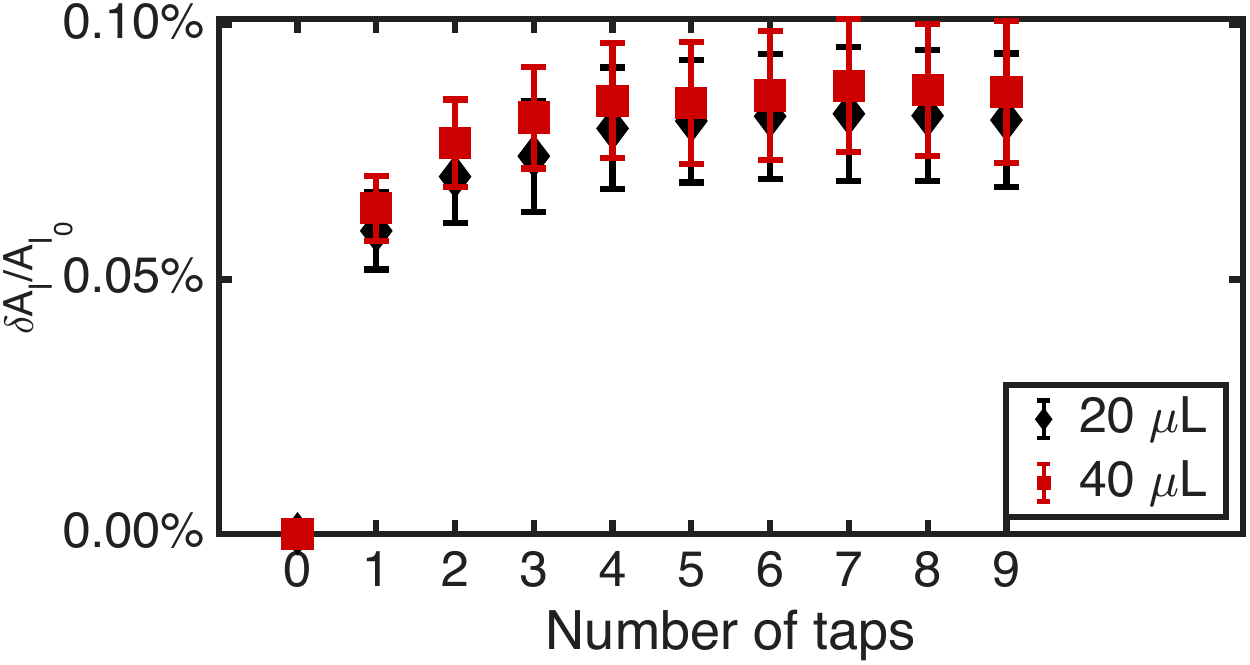}
    \includegraphics[scale=0.29]{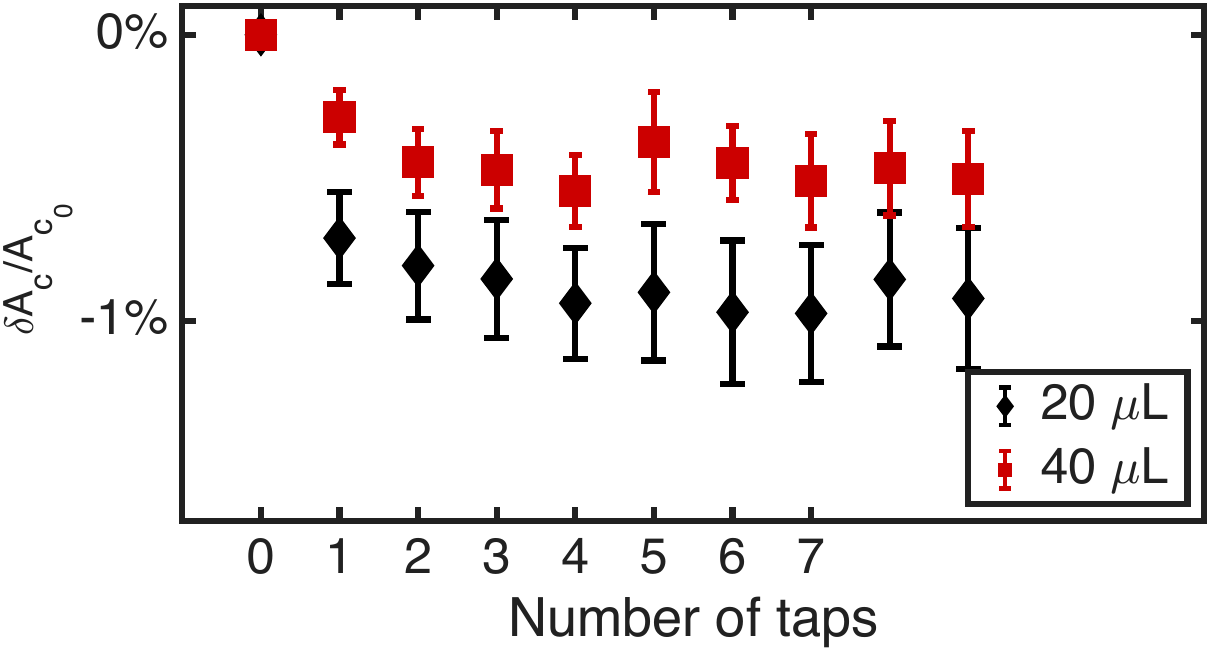}
	\caption{a. Lateral Area vs tapping $20 \mu L$ and $40 \mu L$. b. Contact Area vs tapping $20 \mu L$ and $40 \mu L$. Evaporation corrected in a. and b..}
	\label{Alatcon}
\end{figure*}

After thresholding of the profile the lateral area $A_l$ the contact area with the substrate $A_c$ are measured assuming vertical axisymmetry. To remove evaporation bias and variability between two experiments, We adjust $A_l$ and $A_c$ by a linear law between taps 5 and 10, defining reference lateral area $A_{l0}$ and contact $A_{c0}$ for each experiment. In Fig. \ref{Alatcon} a. we plot lateral area increment $\frac{\delta_{Al}}{A_{l0}}$. After an increase over the three first taps, lateral area increment compared to the drop initial states reaches a plateau. Conversely, contact area increment shows a decrease $\frac{\delta_{Ac}}{A_{c0}}$ over the first few taps before reaching a plateau (Fig. \ref{Alatcon} b.). 

Finally, the total area increment $\frac{\delta_A}{A_{0}}$ shows a slight compaction for $20\mu L$ ($\frac{\delta_A}{A_{0}}=-0.055\% \pm 0.031 \%$) as well as for $40\mu L$ ($\frac{\delta_A}{A_{0}}=0.015 \% \pm 0.026 \%$) but not significant compared to measurement uncertainty. These observations evidence a change of shape of liquid marble due to tapping. In particular, the contact area reduction, being determined by the vertical equilibrium of the droplet, evidence an increase internal pressure.

\begin{figure*}[!h]
	\centering
	\includegraphics[scale=0.29]{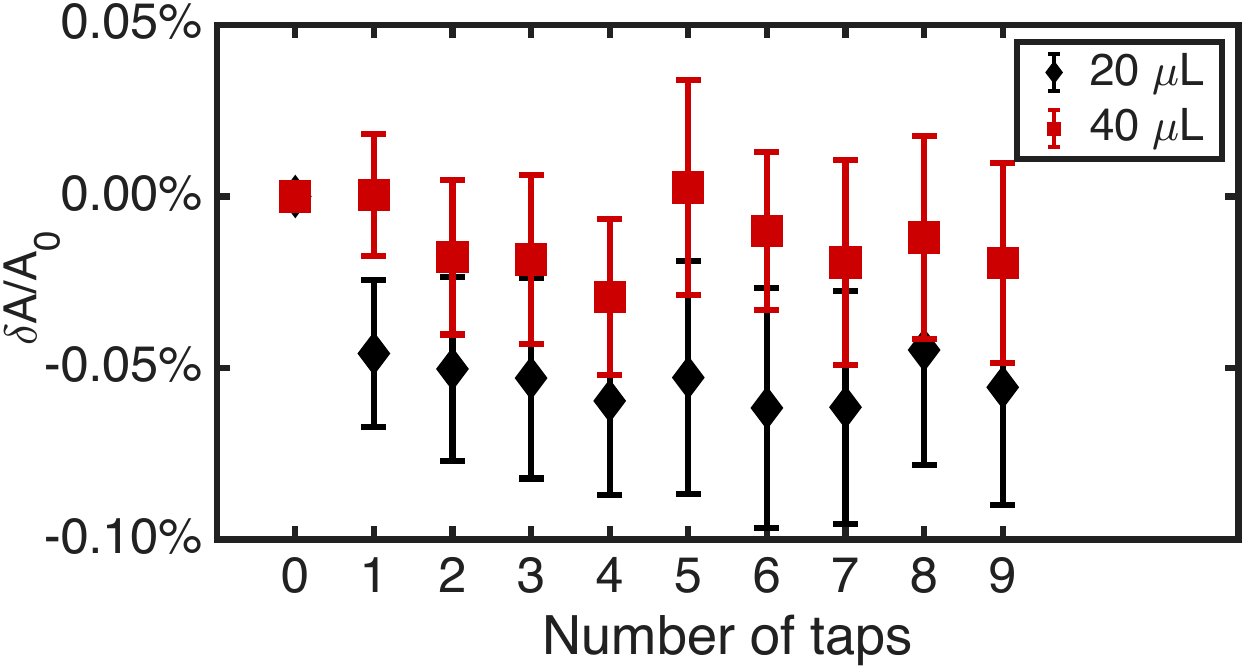}
	\caption{Total Area vs tapping $20 \mu L$ and $40 \mu L$. Evaporation corrected}
	\label{Atot}
\end{figure*}

To go further, we measured the evolution of principal stresses ($\gamma_1$,$\gamma_2$) under tapping using shape analysis: after binarization of the profile image,the local width $ 2 r$ as a function of the distance to the substrate $y$ is extracted and then to find the local principal curvatures adjusted locally over a moving zone of length $500 \mu m$ in curvilinear coordinates, assuming vertical axysimmetry. Then the principal tensions, averaged over the whole drop profile, are estimated using static equilibrium \cite{lagubeau2014armoring}: 
First,the meridian tension $\gamma_1$ is determined only by the vertical equilibrium of the drop above altitude $y$:
$$ M_y g =\pi r^2\Delta P+ 2 \pi r \gamma_1 cos \theta$$
with $y$ the vertical distance from the substrate, $g$ the gravity constant, $M_y$ the partial mass of the drop above $y$, $r$ the width of the drop and $\theta$ the angle of the shell relative to the vertical. Expressing this equilibrium at the substrate level where $\theta =\pi/2$ allowed us to obtain $\delta P$ at the bottom of the drop: $$\delta P = \frac{M_y g}{\pi r^2}$$.
Then the equatorial tension $\gamma_2$ is determined by the local normal equilibrium of the shell: 
$$\Delta P =\gamma_1/R_1 + \gamma_2/R_2 $$
with $\Delta P$ the pressure difference, ($R_1$,$R_2$) the principal curvature radii, and ($\gamma_1$,$\gamma_2$) the associated principal stresses. 

\begin{figure*}[]
	\centering
	\includegraphics[scale=0.29]{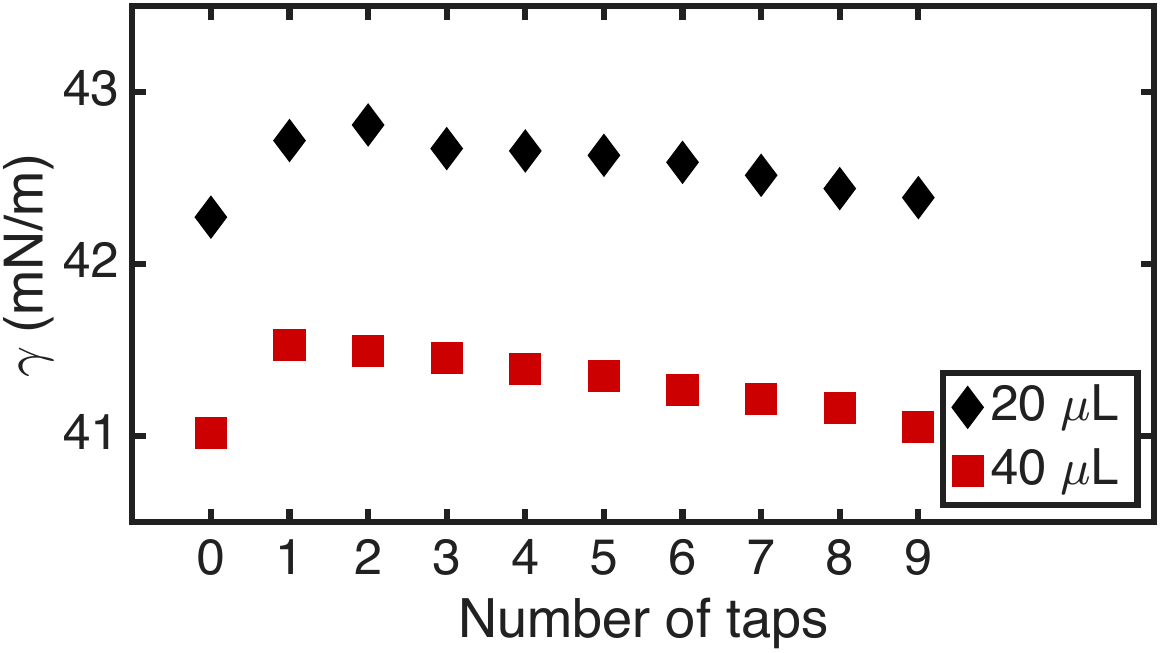}
	\caption{Effective tension vs tapping 20 $\mu L$ and 40 $\mu L$}
	\label{TR}
\end{figure*}

From the best fit, we deduce the principal tensions $ \gamma_1 $, $ \gamma_2 $, the effective surface tension $ \frac{\gamma_1 + \gamma_2}{2} $, and the tension difference $\delta \gamma=\gamma_2-\gamma_1$.

in fig \ref{TR}, we plot the effective surface tension as a function of taps, averaged for all our experiments. We observe an increase of effective surface tension close to $1 mN/m$ after the first tap, and then a slow decrease due to evaporation. Correcting the evaporation as was done for Area measurements, we observe that the surface tension increment as a similar behavior, reaching a plateau after a few taps.

\begin{figure*}[]
	\centering
	\includegraphics[scale=0.29]{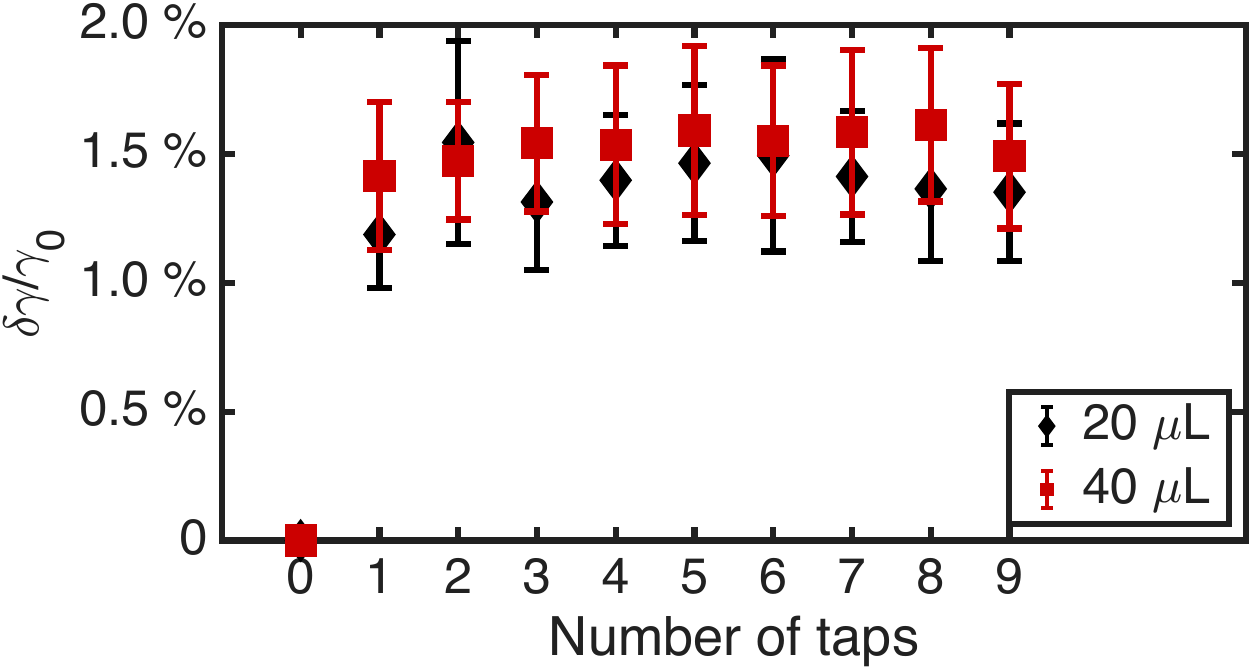}
	\caption{Effective tension increment vs tapping $20 \mu L$ and $40 \mu L$ . Evaporation corrected}
	\label{TN}
\end{figure*}

\begin{figure*}[!h]
	\centering
	\includegraphics[scale=0.29]{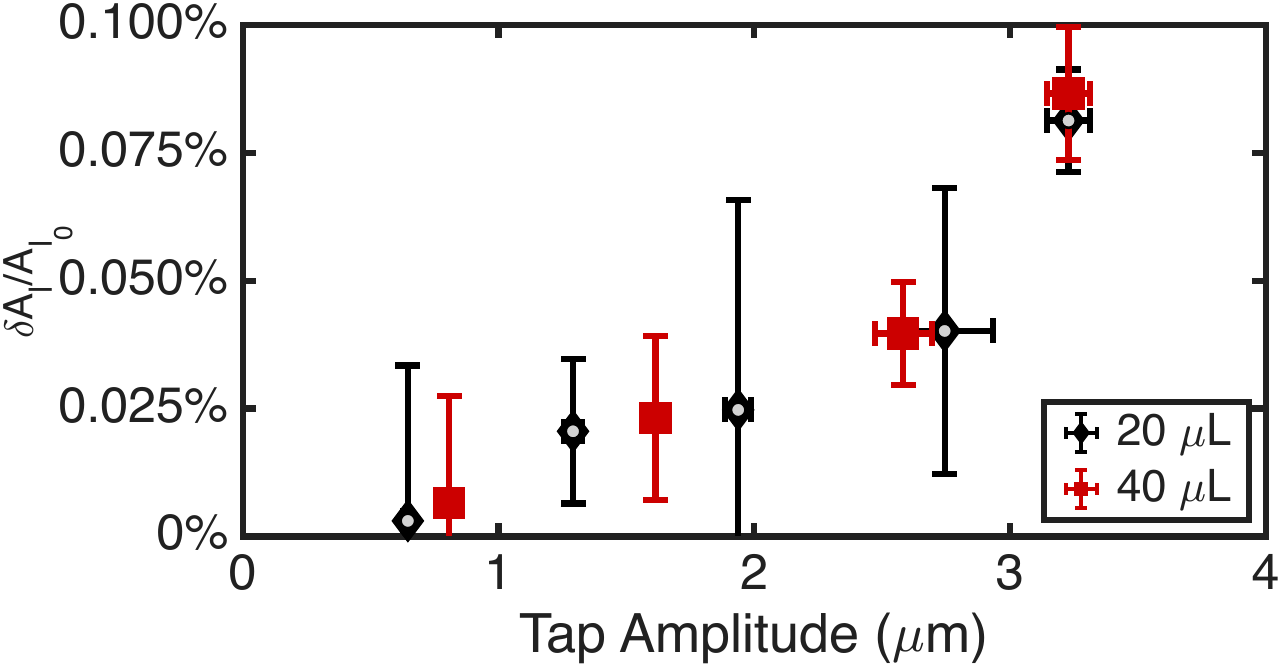}
	\caption{Lateral Area increment vs tapping amplitude $20 \mu L$ and $40 \mu L$.}
	\label{VA}
\end{figure*}

Finally we investigate the effect of amplitude of tap on the drop evolution under tapping. For that purpose, we use the measurements of lower uncertainty, that is the lateral area. We find that, similar to what is observed in granular matter \cite{philippe2002compaction} the more the tap amplitude increases, the more the taps as an effect on the final state (Fig. \ref{VA}).

In conclusion, we evidenced that, similar to dry granular matter tapping, liquid marble properties evolves when submitted to vertical oscillation before reaching a steady state. Their contact area is reduced, inducing an higher internal pressure, their effective surface tension increases, and their surface fraction slightly increases. Additionally, their final state depends on the amplitude of the vertical oscillation. These findings show that a dense liquid marble is usually prepared in a metastable state, and open the possibility of improving control of marbles properties by submitting them to previous vertical taps.
\\
\bibliography{biblio}
\bibliographystyle{unsrt}

\setcounter{figure}{0}
\renewcommand{\thefigure}{S\arabic{figure}}

\section{End Matter}

\textit{Appendix A: Liquid marble preparation ---} The marbles are water droplet of $20 \mu L$ or $40 \mu L$ covered by Lycopodium powder (Aldrich), of diameter $35\pm 7 \mu m$ and bulk density $0.95$. Their preparation protocol is the following: lycopodium powder is deposited on a petri dish on which a drop of water is rolled over a fixed distance. The drop is then rolled on a clean surface to remove excess lycopodium before being placed on the piezoelectric speaker finally allowed to rest for a few tens of seconds. The whole process takes around a minute.
\\

\textit{Appendix B: Experimental determination of the surface fraction ---} Close view camera was used to estimate surface fraction using absorbance of light by lycopodium , of average transmission $\alpha=47\%$ \cite{atkin2011uv}. Knowing the mean gray level $m_g$, illumination gray level $M_g$and  dark background level $d_g$ allow to obtain the fraction of the image occupied by lycopodium on a flatly illuminated image: $\Phi=\frac{M_g-m_g}{(1-\alpha^2)(M_g-d_g)}$.
Indeed, the curvature of the drop distort the illumination as well as contrast, and one need to estimate locally the maximum and minimal envelopes of the illumination. For that purpose we locally estimated the illumination as the maximum gray level (over zones of $177\mu m * 177\mu m$ sufficiently large to image a few grins and meniscus) and then the typical gray level of lycopodium via the maximum of the histogram of gray levels of the zone, allowing a local estimate of dark background level: $d_g=\frac{h_g-\alpha^2.*M_g}{1-\alpha.^2}$. Envelopes of the illumination are finally deduced from both locals estimate maps of $M_g$ and $d_g$ using a second order polynomial fit, and deduce a map of $\phi$ and its average over the whole image $\phi_m$.Finally, the above definition of $\phi$ is only correct on the optical axis, and for a drop shell parallel to the optical plane, inducing a bias to $\phi_m$. Using telecentric lenses, collecting only light parallel to optical axis allows to avoid to have to place accurately center the drop. Slope of the drop shell relative the optical plane induce a variation of $\alpha$, as due to inclination the light  pass inside the shell is longer. For correcting this effect, we adjust  the map of $\phi$ by a second order polynomial whose minimal value correspond on to the undistorted estimation of $\phi$ averaged over the whole image. Correction coefficient was $1.0740 \pm 0.0085$ and $1.0582\pm 0.079$ for $20\mu L$ and $40 \mu L$ drops respectively, with confidence level of $90\%$.
\\

\textit{Appendix C: Imaging---} 
The liquid marble is situated between two backlighted cameras, equipped with telecentric lenses of distinct magnification (Fig.~\ref{Fig1} a.). The full profile is seen at a resolution $19.76 \mu m / pix$ (Fig.~\ref{Fig1} b.), and the local grain network is imaged at a resolution $1.77 \mu m / pix$ (Fig.~\ref{Fig1} c.), covering a fraction of the full area of the droplet ($1.44\%$  and $0.91\%$ for drops of volume $20\mu L$ and $40 \mu L$ respectively).
\\

\textit{Appendix D: tapping signal---} 
 The drop is vibrated vertically using a short burst ("tap") of nine successive frequency steps (from 20 to 100 Hz) of constant amplitude ($3.229 \pm 0.085 \mu m$) sinusoid. Each step is of equal duration ($111 ms$). In the frequency domain, this leads to a nearly flat amplitude between 20 and 100 Hz (Fig.~\ref{Fig1} c.). This frequency range is chosen for efficient energy transfer as it contains the first resonance mode of the drop, and the short duration of the tap is chosen to limit evaporation.
 Profil and local network photos are taken just before the tap. After a delay of $1.15s$, allowing for both relaxation of the droplet and photography, the process is then repeated
\\

\end{document}